\documentclass[conference,a4paper]{IEEEtran}
\usepackage{cite}
\usepackage{epsfig,graphics,mathrsfs,amssymb,amsmath,bm,color,yfonts,txfonts,multirow,slashbox}
\usepackage{subfigure}
\usepackage{algorithm2e}

\begin{document}


\title{Coexistence of Systems with Different Multicarrier Waveforms in LSA Communications}
\author{Mirza Golam Kibria, Gabriel Porto Villardi, Kentaro Ishizu and Fumihide Kojima\\
 Wireless Network Research Institute, National Institute of Information and Communications Technology, Japan\\
 E-mail: $\left\{\text{mirza.kibria, gpvillardi, ishidu, f-kojima}\right\}@\text{nict.go.jp}$ }

\maketitle

\begin{abstract}

We study the coexistence of different multicarrier waveforms such as orthogonal frequency division multiplexing (OFDM), filter-bank multicarrier (FBMC) and universal-filtered multicarrier (UFMC) waveforms in licensed shared access (LSA) for next-generation communication systems. The fundamental changes required in the existing physical layer using OFDM towards a hybrid physical layer (either OFDM-FBMC or OFDM-UFMC) ensuring backward compatibility are discussed.  We also perform mutual interference analysis for the coexisting asynchronous systems sharing the LSA frequency band. Because of the non-orthogonality between their respective transmit signals, power is spilled from a system to the other causing interference. In consideration of analyzing this interaction, power spectral densities of the multicarrier waveforms are exploited. We quantify the amount of percentage power-loss experienced by the interfering systems for not fully exploiting their available power budgets. The simulation results reveal that the interfering system with FBMC suffers the least percentage power-power loss due to its very low side-lobes while conventional OFDM-based system suffers the most. The UFMC-based system exhibits intermediary performance with respect to achieved throughput and power-loss when compared with OFDM and FBMC-based systems.

\end{abstract}

\IEEEpeerreviewmaketitle
\section{Introduction}

Radio spectrum is an extremely valuable resource. The exponential increase in demand for technologies like Wi-Fi or smart electricity grids means we must use this finite resource efficiently. But meeting that growing demand for wireless connectivity is harder in the absence of vacant spectrum. In traditional exclusive licensing systems, many spectrum bands are spatially and temporally underutilized. Due to the scarcity of the spectrum resources and to support the enormous wireless traffic explosion in future, it is important to make full use of the existing radio resources. Spectrum sharing presents a supplementary approach to conventional license-exempt and exclusive licensing schemes. Even though many applications still depend on exclusive access to spectrum, spectrum sharing is increasingly recognized as the breeding ground for wireless innovation that stimulates the development and deployment of more resilient wireless technologies. Licensed shared access (LSA) \cite{Andrews} facilitates dynamic and effective use of available frequency spectrum in next-generation communication networks, and it is likely to become undeniably essential to support the anticipated massive wireless traffic outburst.



The flagship of multicarrier techniques is orthogonal frequency division multiplexing (OFDM). OFDM is currently the transport mechanism base for LTE, LTE-Advanced and WiMAX systems. Although OFDM exhibits several remarkable attributes, it has two inherent shortcomings: susceptiveness to carrier frequency offset (CFO) and a large signal peak-to-average power ratio. With accurate inter-system synchronization, there is no mutual interference between the systems, when different systems use different sets of subcarriers. But if there is synchronization mismatch, the mutual interference will be very high because of the high spectral side-lobes of OFDM signals, which results in low spectral efficiency. At present, the networks of different systems are mostly not synchronized to each other. Furthermore, the OFDM access (OFDMA) based schemes require all systems to use an identical signal frame structure with alignment in time. However, 5G systems need to accommodate a large variety of devices and use cases, which require flexible frame structures. Therefore, the OFDM-based schemes cannot fulfill such 5G requirements.

Finding an appropriate substitute for OFDM is an important issue in 5G research. One basic requirement of 5G is a flexible air interface where the multicarrier attribute like subcarrier spacing is optimizable depending on specific system requirements \cite{Boccardi}. In next-generation systems, devices could communicate based on both open-loop and closed-loop protocols that require parameter optimization, which is probable to be translated into interference. OFDM lacks the ability to address these issues. One of the contender waveforms is filter-bank multicarrier (FBMC)\cite{Schaich}. However, For short burst transmissions, FBMC lacks efficiency due to high time domain overheads \cite{Bellanger}. These shortcomings exhibited by OFDM and FBMC have led some companies and organizations to transfer their focus to finding a more suitable waveform.

A new multicarrier waveform called universal-filtered multicarrier (UFMC)\cite{Schaich}, also referred to as UF-OFDM, has attracted a great deal of attention due to its better efficiency than OFDM. UFMC can be seen as a generalization of OFDM and FBMC. The ultimate goal of employing UFMC is to combine the advantages of OFDM and FBMC while avoiding their main drawbacks. By filtering groups of adjacent subcarriers, the side-lobe levels (compare with OFDM) and the prototype filter length (compare with FBMC) can be simultaneously significantly reduced and becomes adjustable depending on types of applications. UFMC is also found to be a well-suited modulation technique for systems including option for short burst transmissions \cite{Schaich}.

In this paper, we study the coexistence of the above mentioned waveforms in LSA scenario. The fundamental changes required in the existing physical layer using OFDM towards a hybrid physical layer ensuring backward compatibility are discussed. We also perform mutual interference analysis for the coexisting systems sharing the LSA frequency band.


\section{System Model }
\label{SM}

We consider a co-primary LSA communication system with $L$ systems. The LSA spectrum is licensed to multiple systems without defining distinct edge between the bands of different systems. The sharing of the spectrum is accomplished through a common entity called spectrum manager in a coordinated way. The systems have the flexibleness in employing different air-interfaces that support scalable bandwidth and discrete Fourier transnform. A common subcarrier grid of $N_{\rm{c}}$ subcarriers is formed as in \cite{Luo}. Note that the LSA spectrum can be both contiguous and non-contiguous. 


A uniform common subcarrier grid, $\mathcal{U}_{\rm{grid}}$ is assumed, i.e., even if the available LSA spectrum is non-contiguous, $\mathcal{U}_{\rm{grid}}$ covers the available full LSA spectrum. The spectrum assignment among the systems can be performed in two different ways, (i) subcarrier based allocations (ii) fragment based (contiguous band) allocation.
Let ${\{\bm{\mathcal{U}}_1},\cdots, {\bm{\mathcal{U}}_L\}}$  be the sets of indices of the subcarriers assigned to LSA compliant systems and $\bm{\mathcal{U}}_{\rm{grid}}={\bm{\mathcal{U}}_1}\cup{\bm{\mathcal{U}}_2}\cup\cdots\cup{\bm{\mathcal{U}}_L}$. In case of fragment based allocation, the subcarriers in ${\bm{\mathcal{U}}_l}$ are adjacent in $\mathcal{U}_{\rm{grid}}$. In LSA communication, the overall spectrum allocation process is achieved in two stages. In the first stage, the spectrum manager assigns sets of subcarriers or fragments to the systems and in the second stage, the systems perform resources (frequency and power) allocation optimization independently. With perfect synchronization or adequate guard bands between the fragments, each system can transmit data to its own serving users independently without causing interference to the users served by other systems.

\section{Coexistence of OFDM and FBMC/UFMC Waveforms in Hybrid PHY Layer  }

In our waveform coexistence analysis, OFDM is fixed as the waveform of the legacy system and we discuss the fundamental changes required in the existing physical (PHY) mode using only OFDM waveform towards a hybrid PHY mode (either OFDM-FBMC or OFDM-UFMC) ensuring backward compatibility with legacy systems. Note that in 3GPP standardization there is a tendency to reduce options as far as possible for the sake of testing and implementing. We view the hybrid PHY mode as transitory that eases the migration towards the use of either FBMC or UFMC in the future. Therefore, we envisage the use of either FBMC or UFMC in a network such that the stations able to perform FBMC or UFMC should also be able to perform OFDM, at least in the transition period. Note that the stations of the legacy OFDM system do not have the capability to demodulate FBMC/UFMC signals. 

Let us consider a hypothetical intra-system coexistence scenario in the transition period where the BS and some of its associated mobile stations are capable of performing FBMC/UFMC. The association of the FBMC/UFMC stations in the system is achieved by following the normal OFDM procedures such as initial ranging\cite{Sanguinetti}. During the ranging process, the FBMC/UFMC stations notify the BS regarding their capabilities of performing FBMC/UFMC. The BS allocates a certain zones of the frames for FBMC/UFMC transmissions, preferable at the end of the frames, to minimize the disturbance as the stations of the legacy system cannot demodulate FBMC/UFMC signals. Both FBMC and UFMC systems are likely to coexist with OFDM systems. Since FBMC and UFMC are the evolutions of OFDM, some compatibility are expected. All these modulation techniques are based on fast Fourier transform (FFT), and they have a common core. In particular, the initialization phase can be regular to both of the hybrid PHY modes. However, presence of CP in OFDM makes the streaming of the signals different. 
\begin{figure*}
  \centering
   \includegraphics[scale=.065]{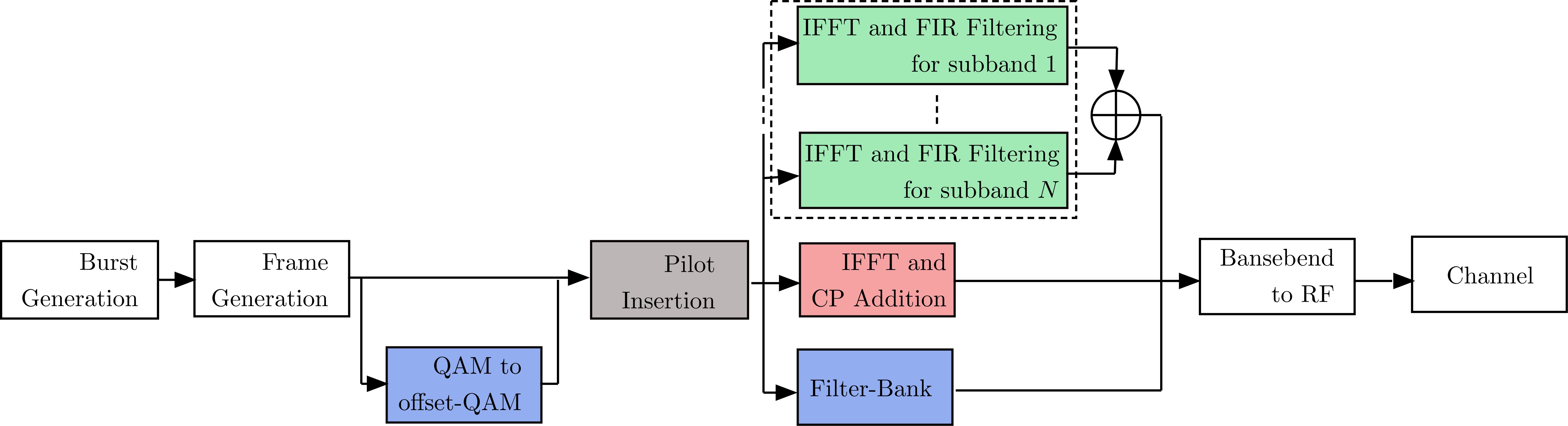}
   \caption{Operation-flow in hybrid PHY mode transmitter including OFDM-FBMC and OFDM-UFMC modes.}
   \label{Trans}
\end{figure*}

\begin{figure*}
  \centering
   \includegraphics[scale=.065]{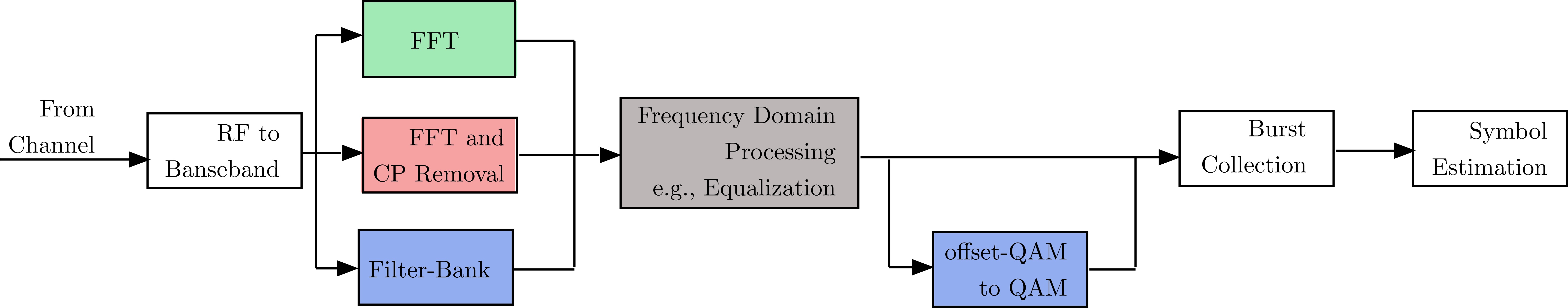}
   \caption{Operation-flow in hybrid PHY mode receiver including OFDM-FBMC and OFDM-UFMC modes.}
   \label{Rec}
\end{figure*}

\subsection{Hybrid PHY mode: OFDM-FBMC}

Compatibility begins at the specifications and system parameters level. The sampling frequency is the same and the subcarrier spacing of FBMC is equal to the subcarrier spacing of OFDM or it is a sub-multiple, so that the FFT is the same, or a subset of the FFT is the same. The difference is that the IFFT is running at double speed for FBMC due to offset-quadrature amplitude modulation (OQAM). In Fig.~\ref{Trans} and Fig.~\ref{Rec}, we illustrate the adjustments required to the legacy system's transmitter to implement a hybrid PHY mode: OFDM-FBMC. Here white blocks are common to both modes while blue blocks are FBMC specific, orange blocks OFDM specific. The gray blocks indicate structures needed in both modes, however, differing in implementation.

Burst generation includes the bit source, forward error coding and symbol mapping. In the multiuser case several bursts are generated. The frame generator builds the frames to be transmitted including the user bursts and the preamble. These two blocks are common between OFDM and FBMC mode. The first adjustment to the legacy OFDM transmitter is performed by adding OQAM block that transforms the QAM to OQAM symbols. This unit is specific to FBMC mode, thus gets bypassed when OFDM mode is active. Pilots are needed in both modes for channel estimation and equalization at the receiver. However, preprocessing of pilot symbols are different. In OFDM mode, pilots are binary phase shift keying (BPSK) symbols. However, in FBMC mode, pilot processing is more complicated compared to OFDM mode. As channel coefficients typically are complex (in baseband notation) the interference caused by adjacent data symbols would disturb channel estimation. To cancel this interference, auxiliary pilot method is used in PHYDYAS. The signal generation in OFDM mode is triggered by inverse FFT (IFFT) and the addition of the CP, while in FBMC mode, the synthesis filter-bank does the job. This is the most significant adjustment needs to be performed to realize hybrid OFDM-FBMC PHY mode.

The signal processing in the hybrid receiver is a bit more complicated, but can be implemented using the same principles followed in transmitter implementation. A large part of the transmitter functions have their counterpart in the receiver. Therefore, the hybrid transmitter design can be reused in the design of a hybrid receiver. The received signal gets transformed employing an FFT after the removal of the CP, if OFDM mode is active. When the FBMC mode is active, the analysis filter-bank performs the transformation. Channel estimation, synchronization and equalization are required in both modes, differing in their implementation. Conversion from OQAM to QAM symbols is needed only in FBMC mode. Like in transmitter, this block gets bypassed if OFDM mode is active. Operations relating to burst collection and symbol estimation/detection are common to both modes.

\subsection{Hybrid PHY mode: OFDM-UFMC}

Note that UFMC is subband-wise filtered and UFMC with filter length 1 is identical to non-CP OFDM. Due to this close relationship, reusing existing OFDM transceiver parts is easy. At the transmitter, all processing leading to complex frequency-domain modulation symbols, e.g. QAM, at the multicarrier modulator input is identical. In Fig.~\ref{Trans} and Fig.~\ref{Rec}, we illustrate the adjustments required to the legacy system's transmitter to implement a hybrid PHY mode: OFDM-UFMC. Here white blocks are common to both modes while green and orange blocks are UFMC and OFDM specific, respectively. The gray blocks indicate structures needed in both modes, however, differing in implementation. Since UFMC also employs QAM, the burst generation, frame generation and pilot insertion blocks are common to both modes. Triggering of signal generation for UFMC mode is a bit different from the OFDM mode. In UFMC mode, for a group of consecutive subcarriers, the IFFT operation is done deriving the corresponding set of IFFT vectors and then filtering by an FIR-filter.

 Regarding the signal processing in the hybrid receiver, when UFMC mode is active, after 2-$N$ FFT, picking each second output and dividing by the frequency response of the filter, yields the same frequency domain scalar per-subcarrier processing as in OFDM. As a result, all existing OFDM channel estimation algorithms can be directly reused by UFMC. In general, just the functionality of multicarrier modulation/demodulation has to be replaced while the rest remains the same.

\section{Mutual Interference Analysis for Coexistence of Systems with Different Multicarrier Waveforms in LSA }

\subsection{Interference Analysis of OFDM Signal }
Assuming an ideal Nyquist pulse, the power density spectrum of any subcarrier in OFDM signal is written as \cite{Weiss}
\begin{equation}
\Psi_{\rm{OFDM}}(f)=PT_s\left(\frac{\sin(\pi fT_s)}{\pi fT_s}\right)^2,
\end{equation}
where $P$ amounts for transmit power on the subcarrier and $T_s$ stands for symbol duration. Therefore, the PDS of any OFDM RB can be expressed as
\begin{equation}
\Phi_{\rm{OFDM}}(f)=\sum_{n=0}^{N_{\rm{RB}}-1}\delta(f-n\Delta f)P_nT_s\left(\frac{\sin(\pi fT_s)}{\pi fT_s}\right)^2,
\end{equation}
Here $N_{\rm{RB}}$ is the number of subcarriers in a RB and $\delta$ is Dirac delta function. $P_n$ is the amount of power allocated to the $n$th subcarrier. This is a sum of shifted, weighted Sinc functions. 
%

\subsection{ Interference Analysis of FBMC Signal }

In an FBMC system, we define the prototype coefficients as $h[l]$, with $l=0,\cdots,L-1$, where $L = KM+1$ and $K$ is the length of each
polyphase components. We define the variable $\mathcal{L} = KM/2$ and assume that the prototype has even symmetry around the $\mathcal{L}$-th coefficient, this means that $h[l] = h[KM-l]$. The PDS of any RB in FBMC signal is given by
%
\begin{equation}
\Phi_{\rm{FBMC}}(f)=\sum_{n=0}^{N_{\rm{RB}}-1}\delta(f-n\Delta f)\overbrace{P_n\left(\sum_{k=-K+1}^{K-1}H_k\frac{\sin(\pi(f-\frac{k}{NK})NK)}{\pi(f-\frac{k}{NK})NK}\right)^2}^{\Psi_{\rm{FBMC}}(f)}\nonumber,
\end{equation}
where $H_k$ are the polyphase components defined in \cite{Bellanger} with $H_0=1$ and $H_k=H_{-k}$. $\Psi_{\rm{FBMC}}(f)$ is the PDS of $n$th subcarrier in the RB. 


\subsection{Interference Analysis of UFMC Signal }

UFMC employs Dolph-Chebyshev\cite{Lynch} filter as an ad hoc choice. One notable property of Chebyshev window 
is that side-lobes attenuation remains same at all frequencies. In UFMC, the original OFDM spectrum is filtered by Chebyshev filter. Thus, the well-known Sinc-spectrum of OFDM also plays a role, which results in the decaying side-lobes in UFMC. The filter coefficients in time-domain are defined as
 \begin{equation}
 \begin{array}{*{35}{l}}
\hspace{-3mm}\Psi_{\rm{UFMC}}^{(n)}=\vspace{1.5mm} \\
\vspace{2mm}
\hspace{-3mm}\left\{\begin{matrix} \left\{ \frac{1}{N}+\frac{(10^{-\alpha/20})}{N}2\sum\limits_{m=1}^{M}{{{C}_{2M}}\left[ {{\kappa }_{0}}\cos \left( \frac{\pi m }{N} \right) \right]\cos \left( \frac{2\pi mn}{N} \right)} \right\},\text{for }\left| n \right|\le M  \nonumber\\
   0,\hspace{65mm}\text{for }\left| n \right|>M  \\
\end{matrix} \right.  \vspace{1.5mm} \\
\end{array}
\end{equation}
where $N=2M+1$ is the UFMC filter length, $\kappa_0$ is a filter parameter and ${C}_{n}$ is the $n$th order Chebyshev polynomial. $\alpha$ defines the side-lobe attenuation in dB. 

\begin{figure}
  \centering
   \includegraphics[scale=.07]{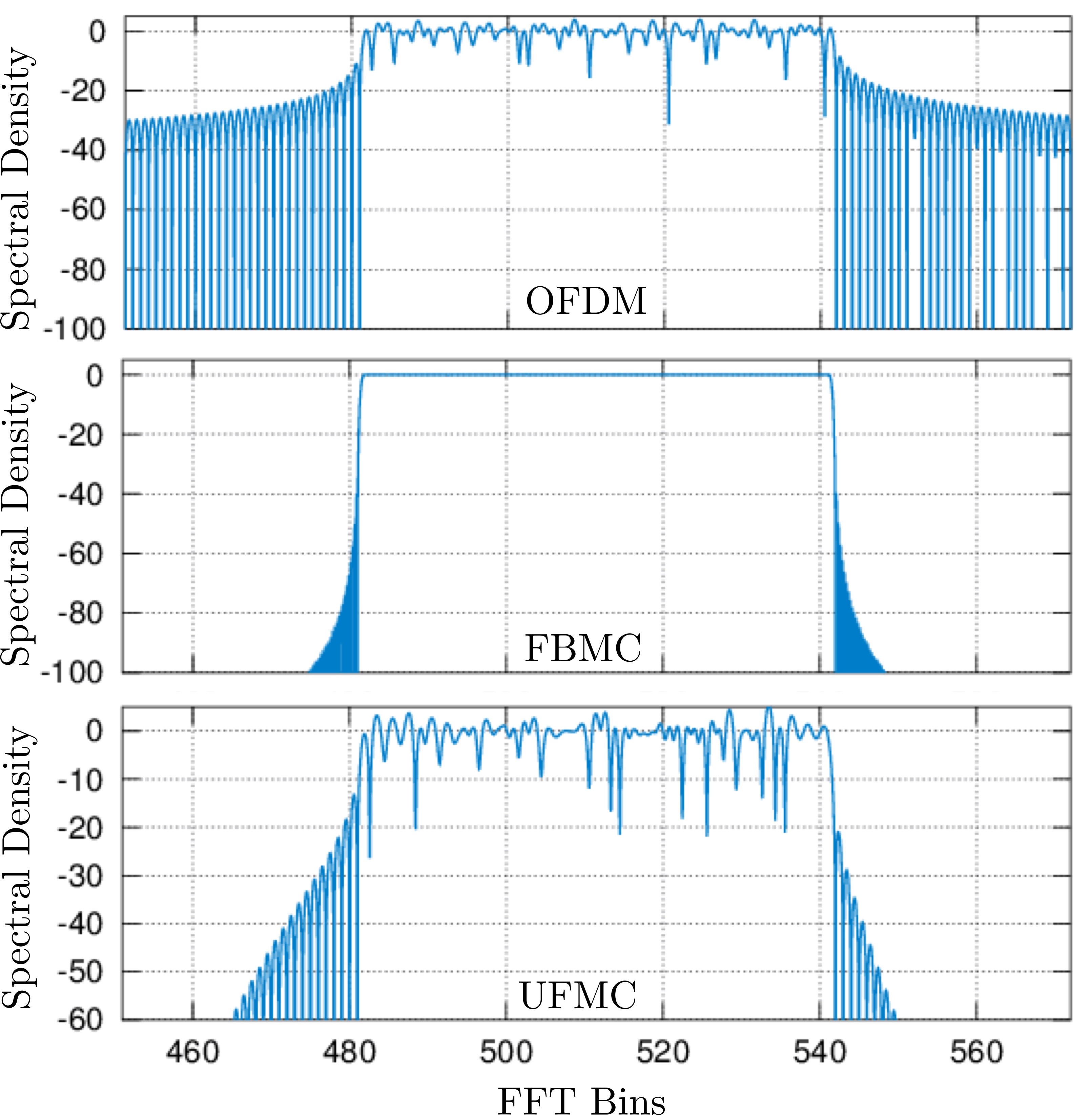}
   \caption{Comparison of power spectral densities with 60 subcarriers. }
   \label{PC9}
\end{figure}

Since UFMC is subband-wise filtered, to generate the PDS of any individual subcarrier of UFMC signal, we send zeros at other subcarrier positions of the RB and the UFMC filter response is shifted to the centre frequency of the RB by multiplying the $l$ the coefficient of the filter with $e^{i2\pi(l-1)f_c\frac{1}{N_{\rm{FFT}}}}$. The amplitude of any subcarrier of UFMC signal in time domain is then obtained by performing convolution between OFDM subcarrier and the centre frequency shifted UFMC filter coefficients. Let the PDS of a single subcarrier in UFMC signal is represented by $\Psi_{\rm{UFMC}}(f )$. Then the 
The PDS of the UFMC RB  is given by
\begin{equation}
\Phi_{\rm{UFMC}}(f)=\sum_{n=0}^{N_{\rm{RB}}-1}\delta(f-n\Delta f)({\Psi_{\rm{UFMC}}(f)})^2.
\end{equation}
Since UFMC is RB-wise filtered, outside of the pass-band (RB width) it has a stronger side-lobe decay than OFDM.
The interference from the interfering system to the other coexisting systems is defined as
 \begin{equation}
  {\rm{I}}_{n}({\rm{d}}_{n})= \int_{({\rm{d}}_{n}-1/2)\Delta f}^{({\rm{d}}_{n}-1/2)\Delta f+B_{\rm{sys}}}\Psi_{\rm{W-Type}}(f)df.
 \end{equation}
Here, ${\rm{d}}_{n}$ is the spectral distance between the $n$th subcarrier of the interfering system and the adjacent system. The total interference introduced to any system is a result of accumulation of interferences from all the RBs available for interfering system's transmission. In Fig.~\ref{PC9}, power spectral densities of systems with different multicarrier waveforms are compared. It clearly shows that FBMC exhibits the best performance in terms of out-of-band emission because of very low side-lobes.
%

\section{Performance Analysis}
\label{PA}
Let us consider an LSA scenario with two coexisting systems operating in contiguous LSA band of 20 MHz bandwidth. We denote the systems as system-A and system-B. The effective bandwidth is taken as 18 MHz with 10\% of the bandwidth assumed to be used for guardband with other nearby systems. Hence, the total number of subcarriers in the subcarrier grid is 1200, with subcarrier spacing of 15 kHz. Consequently, the number of RBs is 100, with 12 subcarriers per RB. The shared 18 MHz spectrum is partitioned into two halves, each for one coexisting system. The coexisting systems may employ different waveforms and obtain orthogonal sets of RBs for transmission. The transmit power budget for each system is taken to be $P_{\rm{max}}$=43 dBm. Each system allocates the resources to its users independently. For simplicity, we consider that each system has one base station (BS) and supports 10 users. The interference-tolerant LSA coexisting systems have their own interference thresholds, e.g. $I_{\rm{th,A}}$ for system-A and $I_{\rm{th,B}}$ for system-B. For simplified performance analysis, we assume that both of the systems have equal threshold values, i.e. $I_{\rm{th,A}}=I_{\rm{th,B}}=I_{\rm{th}}$. We vary the values of $I_{\rm{th}}$ from $10^{-6}$ w to $10^{-1}$ w. For UFMC, the filter length is 74 with $\alpha=40$ dB.

\begin{figure}
  \centering
   \includegraphics[scale=.075]{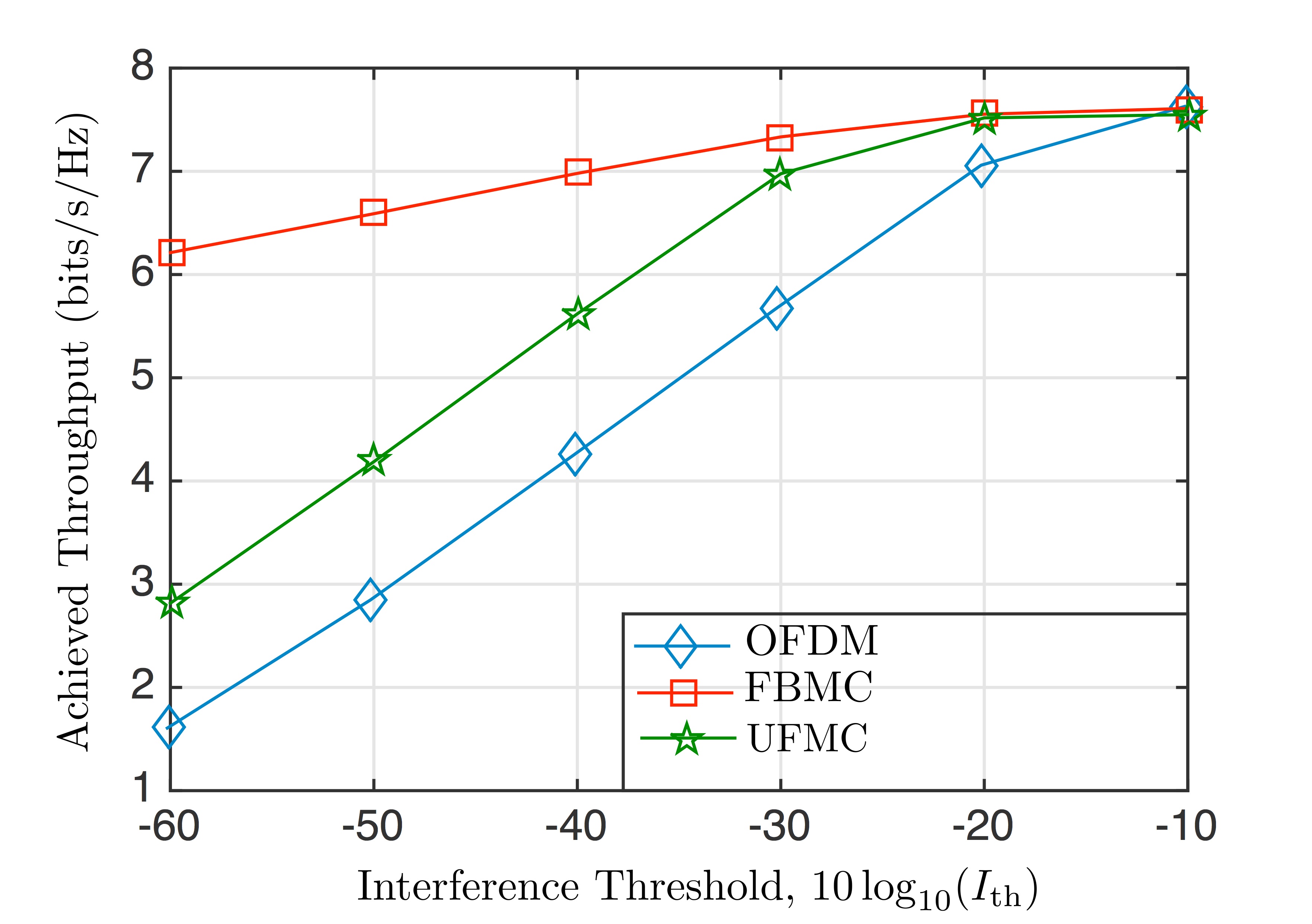}
   \caption{Achieved throughput of system-A vs. interference threshold of system-B (or vice versa).}
   \label{PC2}
\end{figure}
The transmission capacity of the coexisting systems versus interference introduced to adjacent system is plotted for OFDM, FBMC and UFMC in Fig.~\ref{PC2}. It is seen that the achievable throughput of the system can be improved by relaxing the interference threshold of the adjacent systems. OFDM-based system achieves the least downlink capacity compared to FBMC and UFMC based systems while UFMC-based system experiences intermediary performance between OFDM and FBMC-based systems. When the maximum amount of tolerable interference is very small, i.e., for non-robust systems, FBMC exhibits the best performance. Note that the achievable capacity for UFMC based system depends on the chosen value for desired sideband attenuation. Higher the value of sideband attenuation factor $\alpha$, better the achievable downlink capacity. As the interference threshold values increase, therefore, for robust systems, the achievable capacity of FBMC and UFMC-based systems seem to merge depending on the chosen ripple factor and pre-specified interference thresholds. This result motivates the usage of FBMC when no accurate inter-system synchronization is available. Therefore, spectrum sharing are favored scenarios for FBMC, where either the spectrum usage policy is more stringent in terms of emission requirements (such as adjacent channel leakage ratio), or where the requirement is to be more coexistence-friendly to other systems. 

\begin{figure}
  \centering
   \includegraphics[scale=.085]{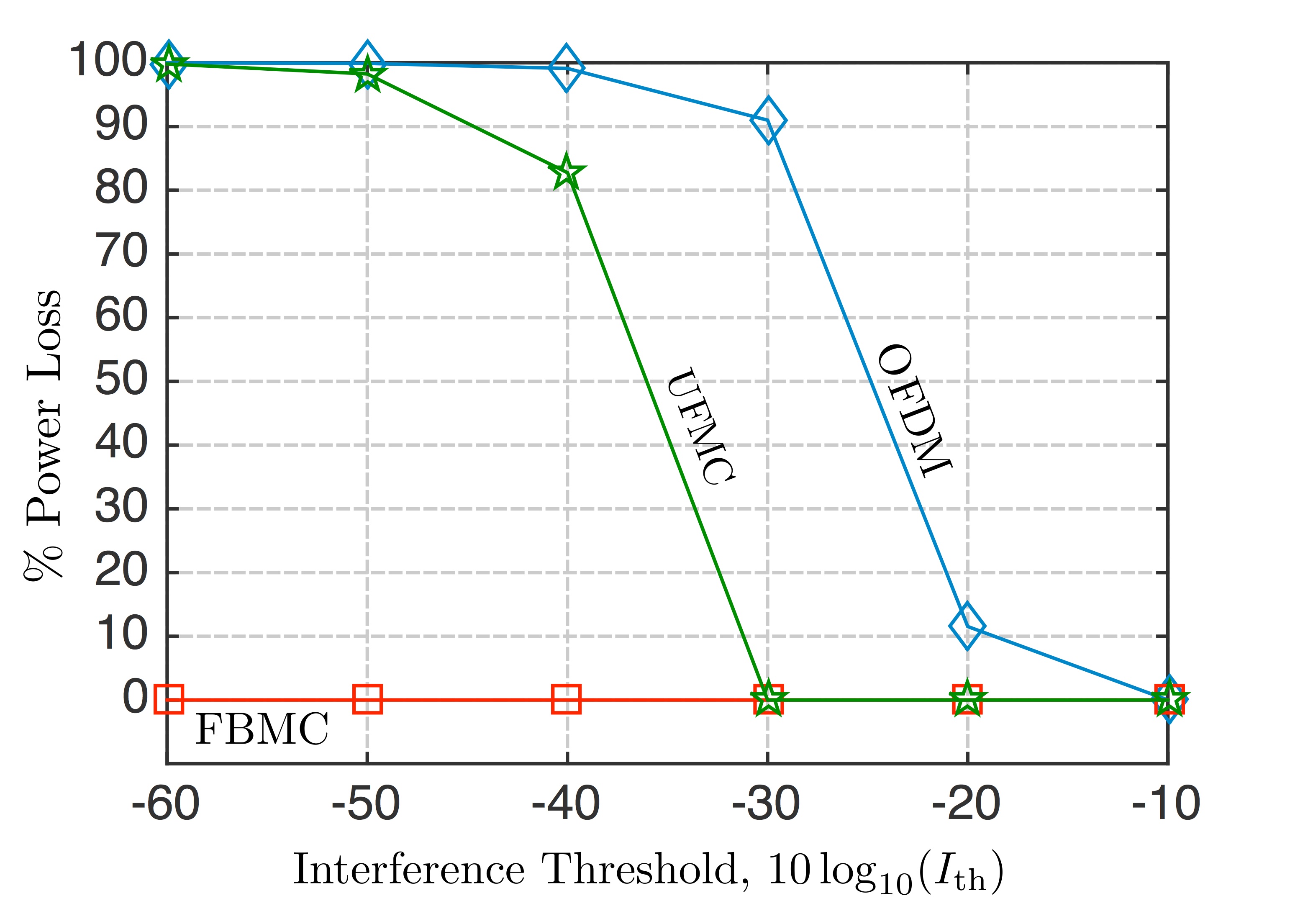}
   \caption{Percentage power-loss of the coexisting systems. The loss is calculated as, $\%\text{power-loss}=100-P_{{\rm{used}},I_{\rm{th}}}/P_{\rm{max}}\times100$, where $P_{{\rm{used}},I_{\rm{th}}}$ is the transmit power used by the the systems to satisfy the interference threshold $I_{\rm{th}}$.}
   \label{PC7}
\end{figure}
In interference-tolerant coexisting systems, a practical way to manage the mutual interference between different systems is by regulating the transmit power in order to comply with the pre-specified interference threshold of different systems in LSA. As a result, the systems cannot fully exploit the maximum benefit out of their available power budget. This is clearly demonstrated in Fig.~\ref{PC7}, where we show the $\%$\textit{power-loss} of the coexisting systems with different waveforms when the pre-specified interference threshold value is varied. It can be observed that OFDM-based system experiences the maximum $\%$\textit{power-loss}, whereas FBMC-based system does not suffer from $\%$\textit{power-loss} at all over the varying $I_{\rm{th}}$ values. On the other hand, UFMC based system suffers much lower $\%$\textit{power-loss} compared to the OFDM-based system. The characteristics of the power-loss curves depend on the spectrum usage policy whether it is more stringent in terms of some emission requirements. For example, when the spectrum usage policy is  not so stringent $(I_{\rm{th}}\in[-30\hspace{1mm} -20])\text{ dBw}$, UFMC-based system does not suffer from any $\%$\textit{power-loss} even with lower $\alpha$. However, for non-robust systems, the $\%$\textit{power-loss} experienced by the UFMC based system increases.

\section{Conclusions}
\label{CC}
In this paper, we investigate the coexistence of different multicarrier waveforms in LSA communications. We discuss the fundamental changes required in the existing physical layer using OFDM waveform towards a hybrid physical layer (either OFDM-FBMC or OFDM-UFMC) ensuring backward compatibility with legacy systems. We have found that hybrid mode terminals can be designed with just a modest increase in complexity of the physical layer compared to a pure OFDM terminal. FBMC and UFMC systems are able to substitute OFDM maintaining a great amount of physical layer compatibility for the future communication networks. We also perform mutual interference analysis for the coexisting asynchronous systems sharing the LSA frequency band. Simulation results have revealed that the interfering system with FBMC suffers the least percentage power power loss due to its very small side-lobes while conventional OFDM-based system suffers the most. The UFMC-based system exhibits intermediary performance with respect to achieved throughput and power loss when compared with OFDM and FBMC-based systems.

\end{document}